\documentclass[twocolumn,showpacs,prb,amsmath,amssymb]{revtex4}
\usepackage{graphicx}

\newcommand{\vvr}{\mathop{\bf r}\nolimits}

\begin{document}
\title{Mean-field Phase Diagram of Two-dimensional Electrons with Disorder in a
Weak Magnetic Field}
\author{Igor S. Burmistrov}
\affiliation{Landau Institute for Theoretical Physics, Kosygina
str. 2, 117940 Moscow, Russia}
\author{Mikhail A. Baranov}
\altaffiliation[Also at  ]{Kurchatov Institute, Kurchatov Square
1, 123182 Moscow, Russia } \affiliation{Institute for Theoretical
Physics, University of Hannover, D-30167, Germany}

\begin{abstract}
We study two-dimensional interacting electrons in a weak
perpendicular magnetic field with the filling factor $\nu \gg 1$
and in the presence of a quenched disorder. In the framework of
the Hartree-Fock approximation, we obtain the mean-field phase
diagram for the partially filled highest Landau level. We find
that the CDW state can exist if the Landau level broadening
$1/2\tau $ does not exceed the critical value
$1/2\tau_{c}=0.038\omega_{H}$. Our analysis of weak
crystallization corrections to the mean-field results shows that
these corrections are of the order of $(1/\nu)^{2/3}\ll 1$ and
therefore can be neglected.
\end{abstract}

 \maketitle


\section{\label{Intro}Introduction}


A two-dimensional electron gas (2DEG) in a perpendicular magnetic
field was a subject of intensive studies, both theoretical and
experimental, for several decades.
The behaviour of the system in a
strong magnetic field where only the lowest Landau level is
occupied, has been investigated in great details~\cite{QHE}. Several
attempts~\cite{MG,BJMA} were made in order to incorporate the case
with larger filling factors $\nu >1$ into the theory. Usually in
these approaches, the ratio of the characteristic Coulomb energy
(at distances of the order of the magnetic length) to the cyclotron
energy, has been assumed to be small.
However, in a weak magnetic field this is not the case, and the
characteristic Coulomb energy exceeds the cyclotron energy. An
attempt to investigate the situation with large Coulomb energy was
made in Ref. \cite{SMG}.

The progress in understanding of the clean 2DEG in a weak magnetic
field was achieved by Aleiner and Glazman~\cite{AG}.
They have derived the low-energy
effective theory on the partially filled highest Landau level by
using the small parameter $1/\nu \ll 1$.
By treating the effective interaction within the Hartree-Fock
approximation, Koulakov, Fogler, and Shklovskii~\cite{KFS}
predicted a unidirectional charge-density-wave (CDW) state (stripe
phase) for the half-filled highest Landau level at zero
temperature and in the absence of disorder.
Moessner and Chalker~\cite{MC} extended the ideas
of Fukuyama, Platzmann and Anderson~\cite{FPA} to the case of a
partially filled highest Landau level and showed
the existence of the mean-field CDW state on the half-filled
Landau level below some temperature $T_{0}$.

Recently, the existence of compressible states near half-filling with
anisotropic transport properties was demonstrated experimentally
for high Landau levels~\cite{LCEPW,DTSPW}. This stimulates an
extensive study of the clean 2DEG in a weak magnetic field and
pinning of stripes by disorder~\cite{F}.

In the clean case, the properties of the CDW states can be
described on the basis of the low energy effective theory for
smooth ``elastic'' deformations~\cite{F}.
Recently, attempts were made to derive such a theory microscopically
starting from the mean-field solution~\cite{Lopat,Fradkin}.
The effects of a quenched disorder on the unidirectional
CDW state (stripe phase) were investigated in the framework of the
phenomenological elasticity theory~\cite{SO}, and a rich variety of
different regimes, which depend on the strength of disorder, were
found. However, to identify the phenomenological parameters of the
theory, a successive  microscopic theory should be developed.

At present, a thorough microscopic analysis of the effects of
disorder on the mean-field transition from the uniform state to
the CDW one, as well as on the phase diagram of the mean-field CDW
states, is absent. The main objective of the present paper is to
investigate these effects on the existence of the mean-field CDW
states in 2DEG in a weak perpendicular magnetic field $H$ (filling
factor $\nu \gg 1$). For the considered case of a large number of
the occupied Landau levels, the mean-field analysis is legitimate
because the fluctuations of the order parameter are strongly
suppressed~\cite{F}. On the other hand, the mean-field approach
cannot be applied to the critical region in the direct vicinity of
the phase transition. This region is however small and does not
lead to any significant uncertainty in our results for critical
temperatures of the transitions.

We assume the presence in the system of a weak quenched disorder,
i.e. the elastic collisions
time satisfies the condition $\tau_{0}\gg \omega_{H}^{-1}$, where
$\omega _{H}=eH/m$ is the cyclotron frequency, $e$ the electron
charge, and $m$ the effective electron mass. Under this condition,
the Landau level broadening $1/2\tau$, which is of the
order of $\sqrt{\omega_{H}\tau _{0}}/\tau _{0}$, is much less than
the spacing $\omega _{H}$ between them. This case can be realized
in high mobility samples which were used for experimental studies
of the anisotropic magnetotransport \cite{LCEPW,DTSPW,inplane}.
Keeping in mind that the relation $T_{0}\sim 1/\tau $ is usually
hold, one expect a much pronounced influence of the quenched
disorder on the properties of electrons on the partially filled
highest Landau level even for a small level broadening $1/2\tau
\ll \omega_{H}$.

One of the main results of our paper is that at zero temperature
the mean-field CDW state is destroyed when the Landau level
broadening exceeds the critical value $1/2\tau_{c}=4 T_{0}/\pi $.
At nonzero temperatures the quenched disorder leads to the
decrease of the temperature of the CDW instability as compared to
the clean case. The physical reason is that the scattering on
impurities breaks the CDW correlations, and therefore results in the
destruction of the coherent CDW state. This is somewhat similar to
the suppression of the critical temperature in conventional
superconductors by magnetic impurities~\cite{AGork} or in
anisotropic superconductors by nonmagnetic impurities~\cite{L}.

The paper is organized as follows. In Sec.~\ref{MFPD.FECDWS} we
introduce the formalism that allows us to evaluate the free energy
of the CDW state in the presence of disorder. In Sec.~\ref{MFPD}
we investigate the instability of the uniform state towards the
formation of the CDW state, and present the mean-field phase
diagrams at the half-filling and arbitrary temperature, and at
zero temperature and arbitrary filling. The weak crystallization
corrections to the mean-field solution are presented in
Sec.~\ref{WCC}. Sec.~\ref{Disc} contains the comparison of the
theory with the recent experimental and numerical results. We end
with conclusions in Sec.~\ref{Conc}.


\section{\label{MFPD.FECDWS}Free energy of the CDW states}


We consider two-dimensional interacting electrons in the presence of a weak
quenched disorder and a weak perpendicular magnetic field. The parameter
that characterizes the strength of the Coulomb interaction is $r_{s}=\sqrt{2}%
e^{2}/\varepsilon v_{F}$ with $v_{F}$ being the Fermi velocity and $%
\varepsilon $ the dielectric constant of a media. We assume that
the Coulomb interaction between the electrons is weak, $r_{s}\ll
1$, and the magnetic field obeys the condition $Nr_{s}\gg 1$,
where $N=[\nu /2]$ is the integer part of $\nu /2$. In this case
it is possible to construct an effective field theory for the
electrons on the highest partially filled Landau level by
integrating out all other degrees of freedom~\cite{AG,B}. We also
assume that the electrons on the partially filled highest Landau
level are spin-polarized. This assumption is based on the
calculations~\cite{BJMA,WS} that shows the existence of fractional
states, composite fermions, and skyrmions only on the lowest and
the first excited Landau levels, as well as on the experimental
observations.

In order to study the transition from the uniform state to the CDW one we
employ the Landau expansion of the free energy in powers of the CDW order
parameter $\Delta (\mathop{\bf q}\nolimits_{j})$, where the vectors $%
\mathop{\bf q}\nolimits_{j}$ that characterize the CDW state, have
the same length~\cite{FPA} $q_{j}=Q$. We perform the expansion up
to the forth order in the CDW order parameter under the assumption
$Nr_{s}^{2}\gg 1$. In this case the Hartree-Fock approximation is
well justified~\cite{MC} because the corrections are small in the
parameter $a_{B}/l_{H}=1/Nr_{s}^{2}\ll 1$,
where $a_{B}=\varepsilon /me^{2}$ is the Bohr radius and $l_{H}=1/\sqrt{%
m\omega _{H}}$ the magnetic length.


\subsection{\label{MFPD.FECDWS.F}Formalism}


The thermodynamical potential of the spin-polarized 2DEG projected
on the $N$th Landau level in the presence of the random potential $V_{dis}(%
\mathop{\bf
r}\nolimits)$ and the magnetic field is given by
\begin{equation}
\Omega =-\frac{T}{N_{r}}\int {\cal D}[\overline{\psi },\psi ]\int {\cal D}%
[V_{dis}]\,{\cal P}[V_{dis}]\,\exp ({\cal S}[\overline{\psi },\psi
,V_{dis}]).  \label{ZStart}
\end{equation}
where the action ${\cal S}[\overline{\psi },\psi ,V_{dis}]$ in the Matsubara
representation has the form
\begin{widetext}
\begin{equation}
{\cal S}=\int_{r}\sum\limits_{\mathop{\omega_{n}}\nolimits}^{\alpha }\Biggl
\{\overline{\psi _{\mathop{\omega_{n}}\nolimits}^{\alpha }}(\mathop{\bf r}%
\nolimits)\Bigl[i\mathop{\omega_{n}}\nolimits+\mu -{\cal H}_{0}-V_{dis}(%
\mathop{\bf r}\nolimits)\Bigr]\psi _{\mathop{\omega_{n}}\nolimits}^{\alpha }(%
\mathop{\bf r}\nolimits)-\frac{T}{2}\sum_{\mathop{\omega_{m}}\nolimits,%
\mathop{\nu_{l}}\nolimits}\int_{\mathop{\bf r}\nolimits^{\prime }}\overline{%
\psi _{\mathop{\omega_{n}}\nolimits}^{\alpha }}(\mathop{\bf r}\nolimits)\psi
_{\mathop{\omega_{n}}\nolimits-\mathop{\nu_{l}}\nolimits}^{\alpha }(%
\mathop{\bf r}\nolimits)U_{0}(\mathop{\bf r}\nolimits,\mathop{\bf r}%
\nolimits^{^{\prime }})\overline{\psi _{\mathop{\omega_{m}}%
\nolimits}^{\alpha }}(\mathop{\bf r}\nolimits)\psi _{\mathop{\omega_{m}}%
\nolimits+\mathop{\nu_{l}}\nolimits}^{\alpha }(\mathop{\bf r}%
\nolimits^{^{\prime }})\Biggr \}.  \label{Sinit}
\end{equation}
\end{widetext}
Here $\psi _{\mathop{\omega_{n}}\nolimits}^{\alpha }(\mathop{\bf r}%
\nolimits) $ and $\overline{\psi _{\mathop{\omega_{n}}\nolimits}^{\alpha }}(%
\mathop{\bf
r}\nolimits)$ are the annihilation and creation operators of an electron on
the $N$th Landau level, $T$ the temperature, $\mu $ the chemical potential, $%
\mathop{\omega_{n}}\nolimits=\pi T(2n+1)$ the Matsubara fermionic frequency,
and $\mathop{\nu_{n}}\nolimits=2\pi Tn$ the bosonic one. The free
Hamiltonian ${\cal H}_{0}$ for 2D electrons with mass $m$ in the
perpendicular magnetic field $H=\epsilon _{ab}\partial _{a}A_{b}$ is ${\cal H%
}_{0}=(-i\nabla -e\vec{A})^{2}/(2m)$. The screened electron-electron
interaction $U_{0}(\mathop{\bf r}\nolimits)$ on the $N$th Landau level takes
into account the effects of interactions with electrons on the other levels,
and has the form (see Ref.~\cite{AG} for the clean case and Ref.~\cite{B}
for the weakly disordered case)
\begin{equation}
U_{0}(q)=\frac{2\pi e^{2}}{\varepsilon q}\frac{1}{\displaystyle1+\frac{2}{%
qa_{B}}\left( 1-\frac{\pi }{6\omega _{H}\tau }\right) \left( 1-{\cal J}%
_{0}^{2}(qR_{c})\right) },  \label{U0}
\end{equation}
where $R_{c}=l_{H}\sqrt{\nu }$ is the cyclotron radius on the $N$th Landau
level and ${\cal J}_{0}(x)$ the Bessel function of the first kind. The range
of the screened electron-electron interaction (\ref{U0}) is determined by
the Bohr radius $a_{B}$. We also assume the Gaussian distribution for the
random potential $V_{dis}(\mathop{\bf r}\nolimits)$
\begin{equation}
{\cal P}[V_{dis}(\vec{r})]=\frac{1}{\sqrt{\pi g}}\,\exp \left( -\frac{1}{g}%
\int_{r}V_{dis}^{2}(\mathop{\bf r}\nolimits)\right) ,  \label{dis}
\end{equation}
where $g=1/\pi \rho \tau _{0}$, $\rho $ is the thermodynamical density of
states, and introduce $N_{r}$ replicated copies of the system labeled by the
replica indices $\alpha =1,...,N_{r}$ in order to average over the disorder.


\subsection{\label{MFPD.FECDWS.HFDAD}The Hartree-Fock decoupling and the average over disorder}


The CDW ground state is characterized by the order parameter $\Delta (%
\mathop{\bf q}\nolimits)$ that is related to the electron density
\begin{equation}
\langle \rho (\mathop{\bf q}\nolimits)\rangle =L_{x}L_{y}n_{L}F_{N}(q)\Delta
(\mathop{\bf q}\nolimits).  \label{rho}
\end{equation}
Here $L_{x}L_{y}$ is the area of the 2DEG, $n_{L}=1/2\pi l_{H}^{2}$ the
number of states on one Landau level, and the form-factor $F_{N}(q)$ is
\begin{equation}
F_{N}(q)=L_{N}\left( \frac{q^{2}l_{H}^{2}}{2}\right) \exp \left( -\frac{%
q^{2}l_{H}^{2}}{4}\right) ,  \label{FF}
\end{equation}
where $L_{N}(x)$ is the Laguerre polynomial. For the case $N\gg 1$, one can
use the following asymptotic expression for the form-factor (\ref{FF})
\begin{equation}
F_{N}(q)={\cal J}_{0}(qR_{c})\qquad ,\qquad qR_{c}\ll \frac{R_{c}^{2}}{%
l_{H}^{2}}=\nu  \label{AFF}
\end{equation}

After the Hartree-Fock decoupling~\cite{HFd} of the interaction term in the
action (\ref{Sinit}) we obtain
\begin{eqnarray}
{\cal S} &=&-\frac{N_{r}\Omega _{\Delta }}{T}+,  \label{SHF1} \\
&+&\int_{r}\sum\limits_{\mathop{\omega_{n}}\nolimits}^{\alpha }\overline{%
\psi _{\mathop{\omega_{n}}\nolimits}^{\alpha }}(\mathop{\bf r}\nolimits)\Bigl%
[i\mathop{\omega_{n}}\nolimits+\mu -{\cal H}_{0}-V_{dis}(\mathop{\bf r}\nolimits%
)+\lambda (\mathop{\bf r}\nolimits)\Bigr]\psi _{\mathop{\omega_{n}}\nolimits%
}^{\alpha }(\mathop{\bf r}\nolimits),  \nonumber
\end{eqnarray}
\begin{equation}
\Omega _{\Delta }=\frac{n_{L}(L_{x}L_{y})^{2}}{2}\int_{q}U(q)\Delta (%
\mathop{\bf q}\nolimits)\Delta (-\mathop{\bf q}\nolimits),  \label{SHF2}
\end{equation}
where the potential $\lambda (\mathop{\bf r}\nolimits)$ results from the
perturbation of the uniform electron density by the charge density wave, and is
connected
with the CDW order parameter as follows
\begin{equation}
\lambda (\mathop{\bf q}\nolimits)=L_{x}L_{y}U(q)F_{N}^{-1}(q)\Delta (
\mathop{\bf q}\nolimits),  \label{lambda}
\end{equation}
and $U(q)=-n_{L}U_{HF}(q)$ with the Hartree-Fock potential $U_{HF}(q)$
given by
\begin{equation}
U_{HF}(q)=U_{0}(q)F_{N}^{2}(q)-\int_{p}\frac{e^{\displaystyle-i\mathop{\bf q}%
\nolimits\mathop{\bf p}\nolimits l_{H}^{2}}}{n_{L}}U_{0}(\mathop{\bf p}%
\nolimits)F_{N}^{2}(\mathop{\bf p}\nolimits).  \label{UHF}
\end{equation}

The averaging over the random potential $V_{dis}(\mathop{\bf r}\nolimits)$
in Eq.(\ref{ZStart}) is straightforward and results in the following quartic
term
\begin{equation}
\frac{g}{2}\int_{r}\sum\limits_{\mathop{\omega_{n}}\nolimits%
\mathop{\omega_{m}}\nolimits}^{\alpha \beta }\overline{\psi _{%
\mathop{\omega_{n}}\nolimits}^{\alpha }}(\mathop{\bf r}\nolimits)\psi _{%
\mathop{\omega_{n}}\nolimits}^{\alpha }(\mathop{\bf r}\nolimits)\overline{%
\psi _{\mathop{\omega_{m}}\nolimits}^{\beta }}(\mathop{\bf r}\nolimits)\psi
_{\mathop{\omega_{m}}\nolimits}^{\beta }(\mathop{\bf r}\nolimits)
\label{quartic2}
\end{equation}
in the action. This term can be decoupled by means of the
Hubbard-Stratonovich transformation ~\cite{HS} with the Hermitian
matrix field variables~\cite{ELK,Finkel} $Q_{nm}^{\alpha \beta
}(\vec{r})$
\begin{equation}
\int {\cal D}[Q]\exp \int_{r}\left[ -\frac{1}{2g}\mathop{\rm tr}\nolimits
Q^{2}(\vec{r})+i\psi ^{\dagger }(\vec{r})Q(\vec{r})\psi (\vec{r})\right] ,
\label{HSQ}
\end{equation}
where the symbol $\mathop{\rm tr}\nolimits$ denotes the matrix trace over
the Matsubara and replica indices. The measure for the functional integral
over the matrix field $Q$ is defined as: the integral (\ref{HSQ}) equals unity
when the fermionic
fields $\psi ^{\dagger }$ and $\psi $ vanish. Note also that in
Eq.(\ref{HSQ}) we introduce the matrix notations according to
\begin{equation}
\psi ^{\dagger }(\cdots )\psi =\sum\limits_{\mathop{\omega_{n}}\nolimits,%
\mathop{\omega_{m}}\nolimits}^{\alpha ,\beta }\overline{\psi _{%
\mathop{\omega_{n}}\nolimits}^{\alpha }}(\cdots )_{nm}^{\alpha \beta }\psi _{%
\mathop{\omega_{m}}\nolimits}^{\beta }.
\end{equation}

After making all these steps, the action becomes
\begin{eqnarray}
{\cal S} &=&-\frac{N_{r}\Omega _{\Delta }}{T}-\frac{1}{2g}\int_{r}%
\mathop{\rm tr}\nolimits Q^{2}+  \nonumber \\
&+&\int_{r}\psi ^{\dagger }(\vec{r})\left( i\omega +\mu -{\cal H}_{0}+\lambda
+iQ\right) \psi (\vec{r}),  \label{S2}
\end{eqnarray}
where $\omega $ is the frequency matrix $(\omega )_{nm}^{\alpha \beta
}=\omega _{n}\delta _{nm}\delta ^{\alpha \beta }$.


\subsection{\label{MFPD.FECDWS.SPQF}The saddle-point in the $Q$ field}


The $Q$ matrix field can be naturally splitted into the transverse $V$ and
the longitudinal $P$ components as follows $Q=V^{-1}PV$. The longitudinal
component $P$ has the block-diagonal structure in the Matsubara space, $%
P_{nm}^{\alpha \beta }\propto \Theta (nm)$, where $\Theta (x)$ is the
Heaviside step function, and corresponds to massive modes. The transverse
component $V$ is a unitary rotation and describes massless (diffusive) modes
(see Refs.~\cite{P1,BPS1} for details).

This decomposition of the variable $Q$ into $P$ and $V$ is motivated by the
saddle-point structure of the action (\ref{S2}) at zero temperature ($\omega
_{n}\rightarrow 0$) and in the absence of the potential $\lambda (%
\mathop{\bf r}\nolimits)$. The corresponding saddle-point solution has the
form $Q_{sp}=V^{-1}P_{sp}V$, where the matrix $P_{sp}$ is
\begin{equation}
(P_{sp})_{nm}^{\alpha \beta }=P_{sp}^{n}\delta _{nm}\delta ^{\alpha \beta }
\label{Pnm}
\end{equation}
with $P_{sp}^{n}$ obeying the equation
\begin{equation}
\pi \rho \tau _{0}P_{sp}^{n}=iG_{0}^{n}(\mathop{\bf r}\nolimits,\mathop{\bf
r}\nolimits).  \label{sp}
\end{equation}
This equation is equivalent to the self-consistent Born approximation
equation~\cite{Ando}. The Green function $G_{0}^{n}(\mathop{\bf r}\nolimits,%
\mathop{\bf r}\nolimits^{^{\prime }})$ is determined as
\begin{equation}
G_{0}^{n}(\mathop{\bf r}\nolimits,\mathop{\bf r}\nolimits^{^{\prime
}})=\sum_{k}\phi _{Nk}^{\ast }(\mathop{\bf r}\nolimits)G_{0}(%
\mathop{\omega_{n}}\nolimits)\phi _{Nk}(\mathop{\bf r}\nolimits^{^{\prime
}}),  \label{G0}
\end{equation}
\begin{equation}
G_{0}(\mathop{\omega_{n}}\nolimits)=[i\mathop{\omega_{n}}\nolimits+\mu
-\epsilon _{N}+iP_{sp}^{n}]^{-1},  \label{Gon}
\end{equation}
where $\epsilon _{N}=\omega _{H}(N+1/2)$ and $\phi _{Nk}(\mathop{\bf r}%
\nolimits)$ are the eigenvalues and eigenfunctions of the hamiltonian
${\cal H}_{0}$, and $k$ denotes pseudomomentum.

In the case of a small disorder $\omega _{H}\tau _{0}\gg 1$ the
solution of equation (\ref{sp}) has the form~\cite{Ando}
\begin{equation}
\begin{array}{lcr}
\displaystyle P_{sp}^{n}=\frac{\mathop{\rm sign}\nolimits\mathop{\omega_{n}}%
\nolimits}{2\tau } & , & \displaystyle\tau =\pi \sqrt{\frac{\rho }{m}}\frac{%
\tau _{0}}{\sqrt{\omega _{H}\tau _{0}}}.
\end{array}
\label{sp1}
\end{equation}

The fluctuations of the $V$ field are responsible for the localization
corrections to the conductivity (in the weak localization regime they
correspond to the maximally crossed diagrams). However, in the considered case,
these corrections are of the order of $1/N\ll 1$ and, therefore, can be
neglected. For this reason we simply put $V=1$.

The presence of the potential $\lambda $ results in a shift of the
saddle-point value (\ref{sp1}) due to the coupling to the fluctuations $%
\delta P=P-P_{sp}$ of the $P$ field. The corresponding effective action for
the $\delta P$ field follows from Eq. (\ref{SHF2}) after
integrating out fermions:
\begin{eqnarray}
{\cal S}[\delta P,\lambda ] &=&\int_{r}\mathop{\rm tr}\nolimits\ln
G_{0}^{-1}-\frac{N_{r}\Omega _{\Delta }}{T}-\frac{1}{2g}\int_{r}\mathop{\rm
tr}\nolimits(P_{sp}+\delta P)^{2}+  \nonumber \\
&+&\int_{r}\mathop{\rm tr}\nolimits\ln \Bigl[1+(i\delta P+\lambda )G_{0}^{-1}%
\Bigr].  \label{SHF3}
\end{eqnarray}
As a result, the thermodynamical potential can be written as
\begin{equation}
\Omega =-\frac{T}{N_{r}}\ln \int {\cal D}[\delta P]I[\delta P]\exp {{\cal S}%
[\delta P,\lambda ],}  \label{Omega}
\end{equation}
where, following Ref.\cite{P1}, the integration measure $I[\delta
P]$ is
\begin{equation}
\ln I[\delta P]=-\frac{1}{(\pi \rho )^{2}}\int \sum\limits_{nm}^{\alpha
\beta }\left[ 1-\Theta (nm)\right] \delta P_{nn}^{\alpha \alpha }\delta
P_{mm}^{\beta \beta }.  \label{mesI}
\end{equation}

The quadratic in $\delta P$ part of the action (\ref{SHF3}) together with
the contribution (\ref{mesI}) from the integration measure determine the
propagator of the $\delta P$ fields (see Ref~\cite{B} for details)
\begin{widetext}
\begin{equation}
\langle \delta P_{m_{1}m_{2}}^{\alpha \beta }(\mathop{\bf q}\nolimits)\delta
P_{m_{3}m_{4}}^{\gamma \delta }(-\mathop{\bf q}\nolimits)\rangle =\frac{g%
\displaystyle\delta _{m_{1}m_{4}}\delta _{m_{2}m_{3}}\delta ^{\alpha \delta
}\delta ^{\beta \gamma }}{\displaystyle1+g\pi _{0}^{m_{1}}(m_{3}-m_{1};q)}-%
\frac{2\left[ 1-\Theta (m_{1}m_{3})\right] }{(\pi \rho )^{2}}\frac{g\delta
_{m_{1}m_{2}}\delta ^{\alpha \beta }}{\displaystyle1+g\pi _{0}^{m_{1}}(0;q)}%
\frac{g\delta _{m_{3}m_{4}}\delta ^{\delta \gamma }}{\displaystyle1+g\pi
_{0}^{m_{3}}(0;q)},  \label{Pcor}
\end{equation}
\end{widetext}
where the bare polarization operator $\pi _{0}^{m}(n;q)$ is
\begin{equation}
\pi _{0}^{m}(n;q)=-n_{L}G_{0}(\mathop{\omega_{m}}\nolimits+\mathop{\nu_{n}}%
\nolimits)G_{0}(\mathop{\omega_{m}}\nolimits)F_{N}^{2}(q).  \label{pi0}
\end{equation}


\subsection{\label{MFPD.FECDWS.TP}Thermodynamical potential}


To find the expansion of the thermodynamical potential $\Omega $ in powers
of the CDW order parameter $\Delta (\mathop{\bf q}\nolimits)$, it is
convenient to introduce a new variable $\widetilde{\delta P}=\delta
P+i\lambda $ and expand $\mathop{\rm tr}\nolimits\ln $ in the action (\ref
{SHF3})\ in powers of this new field $\widetilde{\delta P}$. Then the
thermodynamical potential can be written in the form
\begin{equation}
\Omega =\Omega _{0}+\Omega _{\Delta }+\delta \Omega ,  \label{Omega1}
\end{equation}
where
\begin{equation}
\Omega _{0}(\mu )=\int_{r}\mathop{\rm tr}\nolimits\ln G_{0}^{-1}-\frac{1}{2g}%
\int_{r}\mathop{\rm tr}\nolimits P_{sp}^{2}  \label{Omega0}
\end{equation}
is the mean-field thermodynamical potential of the homogeneous state, and
\begin{equation}
\delta \Omega =-\frac{T}{N_{r}}\ln \int {\cal D}[\widetilde{\delta P}]\exp {%
\tilde{{\cal S}}[\widetilde{\delta P},\lambda ]}  \label{dOmega}
\end{equation}
takes into account the fluctuations of the massive longitudinal field $%
\mathop{\widetilde{\delta P}}\nolimits$ and their interaction with the CDW
order parameter (potential $\lambda $). The action ${\tilde{{\cal S}}[%
\widetilde{\delta P},\lambda ]}$ has the form
\begin{equation}
\tilde{{\cal S}}={\cal S}^{(2)}[\lambda ]+{\cal S}_{int}[\widetilde{\delta P}%
,\lambda ]+{\cal S}^{(2)}[\widetilde{\delta P}]+\sum_{n=3}^{\infty }{\cal S}%
^{(n)}[\mathop{\widetilde{\delta P}}\nolimits],  \label{Stilde}
\end{equation}
with
\begin{equation}
{\cal S}^{(2)}[\lambda ]=\frac{N_{r}}{2g}\sum_{\mathop{\omega_{n}}\nolimits%
}\int_{r}\lambda (\mathop{\bf r}\nolimits)\lambda (\mathop{\bf r}\nolimits),
\label{S2lambda}
\end{equation}
\begin{equation}
{\cal S}_{int}[\widetilde{\delta P},\lambda ]=-\frac{i}{g}\int_{r}\lambda (%
\mathop{\bf r}\nolimits)\mathop{\rm tr}\nolimits\mathop{\widetilde{\delta P}}%
\nolimits(\mathop{\bf r}\nolimits),  \label{Sint}
\end{equation}
and
\begin{equation}
{\cal S}^{(n)}[\mathop{\widetilde{\delta P}}\nolimits]=\frac{(-i)^{n}}{n}%
\mathop{\rm tr}\nolimits\prod_{j=1}^{n}\int_{r_{j}}\mathop{\widetilde{\delta
P}}\nolimits(\mathop{\bf r}\nolimits_{j})G_{0}(\mathop{\bf r}\nolimits_{j}%
\mathop{\bf r}\nolimits_{j+1}),  \label{Sn}
\end{equation}
where $\mathop{\bf r}\nolimits_{n+1}=\mathop{\bf r}\nolimits_{1}$. Note that
the terms in the action (\ref{Stilde}), that are proportional to $N_{r}^{2}$%
, are omitted because they do not contribute to $\delta \Omega $ in the
replica limit $N_{r}\rightarrow 0$. Another important observation is that
the propagator of the $\mathop{\widetilde{\delta P}}\nolimits$ fields is the
same as for the $\delta P$ fields (\ref{Pcor}).

By using Eqs.(\ref{dOmega})-(\ref{Sn}) we can write
\begin{equation}
\delta \Omega =-\frac{T}{2g}\sum_{\mathop{\omega_{n}}\nolimits%
}\int_{r}\lambda (\mathop{\bf r}\nolimits)\lambda (\mathop{\bf r}\nolimits)-%
\frac{T}{N_{r}}\ln \left\langle \exp {\tilde{{\cal S}}_{int}}\right\rangle ,
\label{dOmega1}
\end{equation}
where $\langle \cdots \rangle $ denotes the average over $%
\mathop{\widetilde{\delta P}}\nolimits$ with respect to the action $\tilde{%
{\cal S}}[\mathop{\widetilde{\delta P}}\nolimits,0]$. This equation allows
us to find the contributions to the thermodynamical potential $\Omega $ up
to any order of the CDW order parameter $\Delta (\mathop{\bf q}\nolimits%
)=F_{N}(q)U(q)^{-1}\lambda (\mathop{\bf q}\nolimits)/L_{x}L_{y}$.

In this paper we will work only with the expansion upto the fourth
order term (the Landau expansion). This implies that our
consideration is valid only close to the transition point where
the value of the order parameter is small and one can truncate the
series (\ref{dOmega}) after several first terms. It should be
mentioned, however, that we should avoid a direct vicinity of the
phase transition (the critical region, for more details see
Sec.~\ref{WCC}) where the fluctuations of the order parameter
break the mean-field approach.


\subsubsection{\label{MFPD.FECDWS.TP.SOC}Second-order contribution}


The  second order contribution to the thermodynamical potential $\delta
\Omega $ is
\begin{equation}
\delta \Omega ^{(2)}=-\frac{T}{2g}\sum_{\mathop{\omega_{n}}\nolimits%
}\int_{r}\lambda (\mathop{\bf r}\nolimits)\lambda (\mathop{\bf r}\nolimits)-%
\frac{T}{2N_{r}}\left\langle S_{int}^{2}\right\rangle _{0},  \label{so1}
\end{equation}
where $\langle \cdots \rangle _{0}$ stands for the average over $%
\mathop{\widetilde{\delta P}}\nolimits$ with respect to the action $\tilde{%
{\cal S}}^{(2)}[\mathop{\widetilde{\delta P}}\nolimits]$. We replace the
average over the full action $\tilde{{\cal S}}[\mathop{\widetilde{\delta P}}%
\nolimits,0]$ by the average over the quadratic part $\tilde{{\cal S}}^{(2)}[%
\mathop{\widetilde{\delta P}}\nolimits]$ only because the higher order in $%
\mathop{\widetilde{\delta P}}\nolimits$ terms lead to the contributions that
are proportional to $N_{r}^{2}$, and therefore vanish in the replica limit $%
N_{r}\rightarrow 0$.

\begin{figure}
\includegraphics[width=150pt,height=70pt]{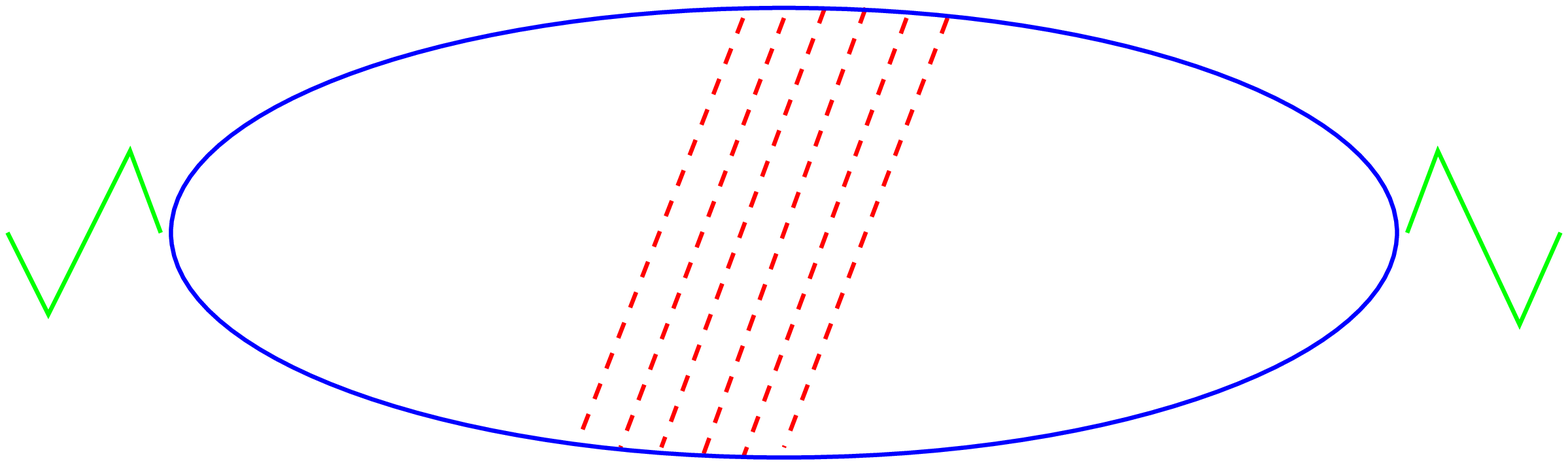}
\caption{ Second-order contribution to the thermodynamic
potential. Solid line denotes electron Green function, dashes are
impurity lines and vertexes are $\protect\lambda (\mathop{\bf
r}\nolimits)$} \label{FIG.MFPD.FECDWS.SOC}
\end{figure}

With the help of Eqs.(\ref{lambda}),(\ref{Pcor}), and (\ref{Sint}), we obtain
\begin{equation}
\frac{\displaystyle\delta \Omega ^{(2)}}{L_{x}^{2}L_{y}^{2}}=n_{L}\frac{T}{2}%
\sum_{\mathop{\omega_{n}}\nolimits}\int_{q}\frac{\displaystyle %
U^{2}(q)G_{0}^{2}(\mathop{\omega_{n}}\nolimits)}{\displaystyle1+g\pi _{0}^{%
\mathop{\omega_{n}}\nolimits}(0,q)}\Delta (\mathop{\bf q}\nolimits)\Delta (-%
\mathop{\bf q}\nolimits).  \label{so2}
\end{equation}
The corresponding diagram in the usual ``cross technique'' is
shown in Fig.~\ref{FIG.MFPD.FECDWS.SOC}.


\subsubsection{\label{MFPD.FECDWS.TP.TOC}Third-order contribution}

\begin{figure}
\includegraphics[width=150pt,height=70pt]{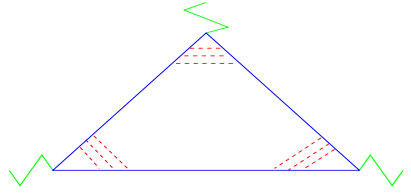}
\label{MFPD.FECDWS.TP.TOC} \caption{ The third-order contribution
to the thermodynamical potential.} \label{FIG.MFPD.FECDWS.TOC}
\end{figure}

The contribution of the third power of the CDW order parameter to the
thermodynamical potential $\delta \Omega ^{(3)}$ can be written as
\begin{equation}
\delta \Omega ^{(3)}=-\frac{T}{3!N_{r}}\left\langle S_{int}^{3}\right\rangle
_{\mathop{\widetilde{\delta P}}\nolimits}^{(c)}=-\frac{T}{3!N_{r}}%
\left\langle S_{int}^{3}S^{(3)}\right\rangle _{0}^{(c)},  \label{to1}
\end{equation}
where the superscript $(c)$ indicates that only connected diagrams are taken
into account. Here we omit again the terms that vanish in the replica limit $%
N_{r}\rightarrow 0$. After performing the averaging over $%
\mathop{\widetilde{\delta P}}\nolimits$ with the help of Eqs.(\ref{lambda}),(%
\ref{Pcor}),(\ref{Sint}), and (\ref{Sn}), we obtain
\begin{eqnarray}
\frac{\delta \Omega ^{(3)}}{L_{x}^{3}L_{y}^{3}} &=&(2\pi )^{2}n_{L}\frac{T}{3}%
\sum_{\mathop{\omega_{n}}\nolimits}\prod_{j=1}^{3}\Biggl [\int_{q_{j}}\frac{%
U(q_{j})\Delta (\mathop{\bf q}\nolimits_{j})G_{0}(\mathop{\omega_{n}}%
\nolimits)}{\displaystyle1+g\pi _{0}^{\mathop{\omega_{n}}\nolimits}(0,q_{j})}%
\Biggr ]  \nonumber \\
&\times &\delta (\mathop{\bf q}\nolimits_{1}+\mathop{\bf q}\nolimits_{2}+%
\mathop{\bf q}\nolimits_{3})\exp \frac{i}{2}\Bigl%
(q_{1}^{x}q_{2}^{y}-q_{1}^{y}q_{2}^{x}\Bigr).  \label{to2}
\end{eqnarray}
The contribution $\delta \Omega ^{(3)}$ corresponds to the diagram
in Fig.~\ref{FIG.MFPD.FECDWS.TOC}.


\subsubsection{\label{MFPD.FECDWS.TP.FOC}Forth-order contribution}

\begin{figure}
\includegraphics[width=150pt,height=70pt]{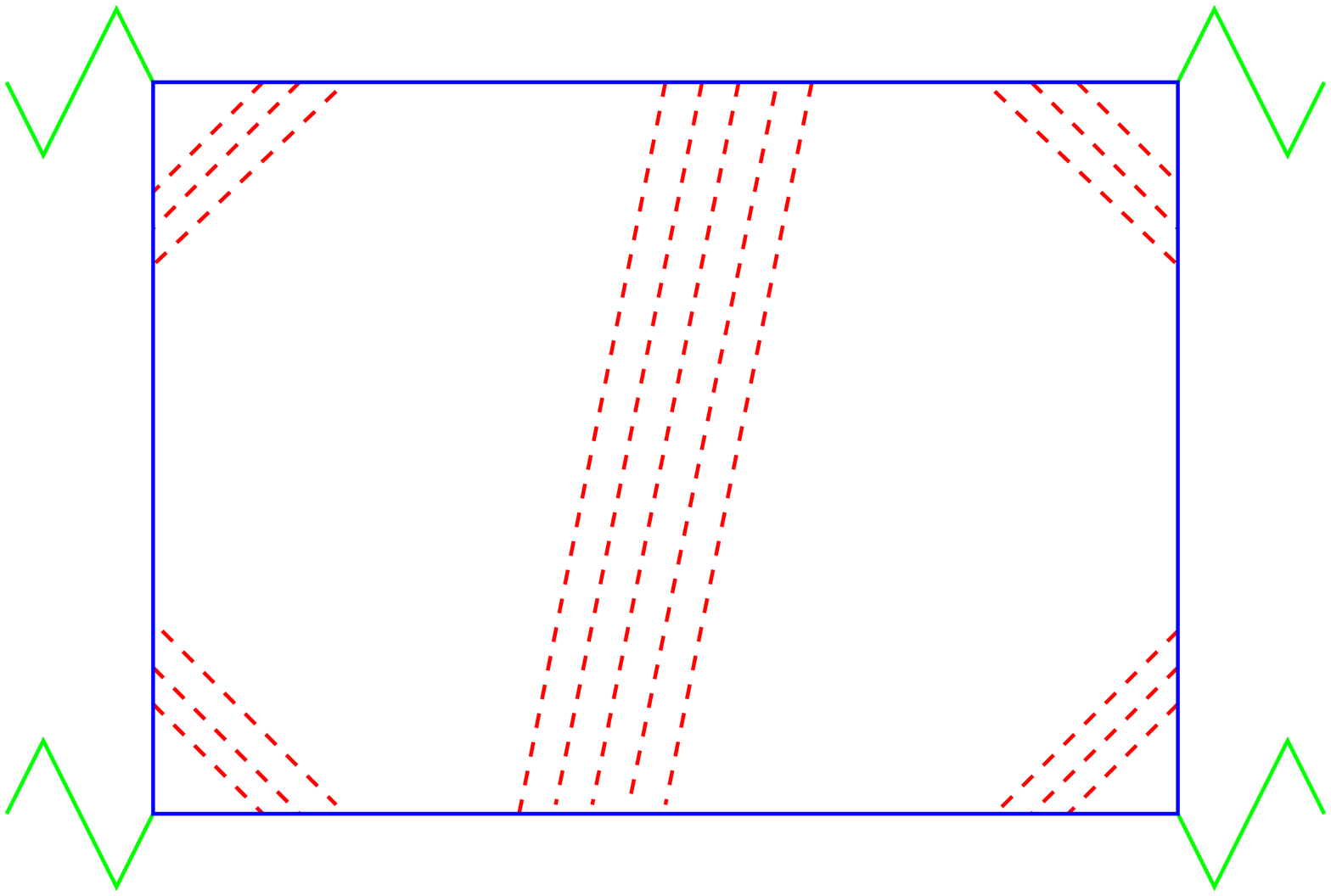}
\caption{ The forth-order contribution to the thermodynamical potential.}
\label{FIG.MFPD.FECDWS.FOC}
\end{figure}

The forth order contribution $\delta \Omega ^{(4)}$ is
\begin{eqnarray}
\delta \Omega ^{(4)} &=&-\frac{T}{4!N_{r}}\left\langle
S_{int}^{4}\right\rangle ^{(c)}  \nonumber \\
&=&-\frac{T}{4!N_{r}}\left\langle S_{int}^{4}\Bigl[S^{(4)}+\frac{1}{2}%
(S^{(3)})^{2}\Bigr]\right\rangle _{0}^{(c)},  \label{fo1}
\end{eqnarray}
where again only terms which is proportional to $N_{r}$ are kept. By using
Eqs.(\ref{lambda}),(\ref{Pcor}),(\ref{Sint}), and (\ref{Sn}), we find
\begin{eqnarray}
\frac{\delta \Omega ^{(4)}}{L_{x}^{4}L_{y}^{4}} &=&(2\pi )^{2}n_{L}\frac{T}{4%
}\sum_{\mathop{\omega_{n}}\nolimits}\prod_{j=1}^{4}\Biggl [\int_{q_{j}}\frac{%
U(q_{j})\Delta (\mathop{\bf q}\nolimits_{j})G_{0}(\mathop{\omega_{n}}%
\nolimits)}{\displaystyle1+g\pi _{0}^{\mathop{\omega_{n}}\nolimits}(0,q_{j})}%
\Biggr ]  \nonumber \\
&\times &\delta (\mathop{\bf q}\nolimits_{1}+\mathop{\bf q}\nolimits_{2}+%
\mathop{\bf q}\nolimits_{3}+\mathop{\bf q}\nolimits_{4})\frac{\displaystyle%
1-g\pi _{0}^{\mathop{\omega_{n}}\nolimits}(0,|\mathop{\bf q}\nolimits_{1}+%
\mathop{\bf q}\nolimits_{2}|)}{\displaystyle1+g\pi _{0}^{\mathop{\omega_{n}}%
\nolimits}(0,|\mathop{\bf q}\nolimits_{1}+\mathop{\bf q}\nolimits_{2}|)}
\nonumber \\
&\times &\exp \frac{i}{2}\Bigl(q_{1}^{x}q_{2}^{y}-q_{1}^{y}q_{2}^{x}\Bigr%
)\exp \frac{i}{2}\Bigl(q_{3}^{x}q_{4}^{y}-q_{3}^{y}q_{4}^{x}\Bigr).
\label{fo2}
\end{eqnarray}
In the usual ``cross technique'' the contribution $\delta \Omega
^{(4)}$ corresponds  to the diagram shown in
Fig.~\ref{FIG.MFPD.FECDWS.FOC}.


\subsection{\label{MFPD.FECDWS.FE}Free energy}


The free energy of the CDW state can be written in the form
\begin{equation}
{\cal F}={\cal F}_{0}+\Omega (\mu )-\Omega _{0}(\mu _{0})+(\mu -\mu
_{0})N_{e},  \label{FE1}
\end{equation}
where ${\cal F}_{0}$ is the free energy of the normal (homogeneous) state, $%
N_{e}$ the total number of electrons, $\mu $ and $\mu _{0}$ the chemical
potentials of the CDW state and the normal state respectively.

In order to find the free energy of the CDW state to the forth order in the
CDW order parameter we expand $\Omega _{0}(\mu _{0})$ around the point $\mu $
to the second order in $\mu -\mu _{0}$. This results in
\begin{equation}
{\cal F}={\cal F}_{0}+\Omega (\mu )-\Omega _{0}(\mu )-\frac{1}{2}(\mu -\mu
_{0})^{2}\frac{\partial ^{2}\Omega _{0}}{\partial ^{2}\mu _{0}}.  \label{FE2}
\end{equation}
The difference $\mu -\mu _{0}$ of the chemical potentials in the CDW and the
normal states is
\begin{equation}
\mu -\mu _{0}=\frac{\displaystyle\partial \delta \Omega }{\displaystyle%
\partial N_{e}}\left( \frac{\displaystyle\partial N_{e}}{\displaystyle%
\partial \mu }\right) ^{-1},  \label{dmu}
\end{equation}
and from Eq.(\ref{Omega1}) we obtain
\begin{equation}
{\cal F}={\cal F}_{0}+\Omega _{\Delta }+\delta \Omega +\frac{1}{2}\left(
\frac{\partial \delta \Omega ^{(2)}}{\partial \mu }\right) ^{2}\left( \frac{%
\partial N_{e}}{\partial \mu }\right) ^{-1}.  \label{FE3}
\end{equation}
With the expression (\ref{so2}) for $\delta \Omega ^{(2)}$ this gives
\begin{widetext}
\begin{equation}
{\cal F}={\cal F}_{0}+\Omega _{\Delta }+\delta \Omega +\frac{%
n_{L}(L_{x}L_{y})^{3}}{2}\Biggl [T\sum_{\mathop{\omega_{n}}\nolimits}\int_{q}%
\frac{\displaystyle U^{2}(q)G_{0}^{3}(\mathop{\omega_{n}}\nolimits)}{[%
\displaystyle1+g\pi _{0}^{\mathop{\omega_{n}}\nolimits}(0,q)]^{2}}\Delta (%
\mathop{\bf q}\nolimits)\Delta (-\mathop{\bf q}\nolimits)\Biggr ]^{2}\Bigl%
[T\sum_{\mathop{\omega_{n}}\nolimits}G_{0}^{2}(\mathop{\omega_{n}}\nolimits)%
\Bigr]^{-1}.  \label{FE4}
\end{equation}
\end{widetext}


\subsection{\label{MFPD.FECDWS.FETCDWS}Free energy of the triangular CDW state}


The CDW order parameter for the triangular lattice symmetry (bubble phase)
can be written in the form~\cite{FPA}
\begin{equation}
\Delta (\mathop{\bf q}\nolimits)=\frac{(2\pi )^{2}}{L_{x}L_{y}}\Delta
(Q)\sum_{j=1}^{3}\Bigl[\delta (\mathop{\bf q}\nolimits-\mathop{\bf Q}%
\nolimits_{j})+\delta (\mathop{\bf q}\nolimits+\mathop{\bf Q}\nolimits_{j})%
\Bigr],  \label{DeltaT}
\end{equation}
where the vectors $\mathop{\bf Q}\nolimits_{j}$ have the angle $2\pi /3$
between each other and obey the condition $\mathop{\bf Q}\nolimits_{1}+%
\mathop{\bf Q}\nolimits_{2}+\mathop{\bf Q}\nolimits_{3}=0$.

By using Eqs.(\ref{so2}),(\ref{to2}),(\ref{fo2}) and (\ref{FE4}), we obtain
the following expression for the free energy of the triangular CDW state
\begin{equation}
{\cal F}^{t}={\cal F}_{0}+4 \frac{L_{x}L_{y}}{2\pi
l_{H}^{2}}T_{0}(Q)\Bigl[a_{2}\Delta ^{2}+a_{3}\Delta
^{3}+a_{4}\Delta ^{4}\Bigr].  \label{FETri}
\end{equation}
Here the three coefficients $a_{1}$, $a_{2}$, and $a_{3}$ of the Landau
expansion are as follows
\begin{equation}
a_{2}=3\Bigl[1-\frac{T_{0}(Q)}{\pi ^{2}T}\sum_{n}\frac{1}{\xi
_{n}^{2}+\gamma ^{2}(Q)}\Bigr],  \label{a2}
\end{equation}
where
\begin{equation}
\xi _{n}=n+\frac{1}{2}+\frac{1}{4\pi T\tau }-i\,\frac{\mu }{2\pi T}%
\,,\,\gamma (Q)=\frac{F_{N}(Q)}{4\pi T\tau }  \label{xin}
\end{equation}
and $T_{0}(Q)=U(Q)/4$,
\begin{equation}
a_{3}=i\,8\frac{T_{0}^{2}(Q)}{\pi ^{3}T^{2}}\cos \left( \frac{\sqrt{3}Q^{2}}{%
4}\right) \sum_{n}\frac{\xi _{n}^{3}}{\Bigl[\xi _{n}^{2}+\gamma ^{2}(Q)\Bigr%
]^{3}},  \label{a3}
\end{equation}
and
\begin{widetext}
\begin{eqnarray}
a_{4} &=&\frac{24T_{0}^{3}(Q)}{\pi ^{4}T^{3}}\Biggl \{\frac{1}{2}\sum_{n}%
\frac{\xi _{n}^{4}}{\Bigl[\xi _{n}^{2}+\gamma ^{2}(Q)\Bigr]^{4}}\Biggl [%
3D_{n}(0)+\Bigl(1+\cos \frac{\sqrt{3}Q^{2}}{2}\Bigr)\Bigl(D_{n}(Q)+D_{n}(%
\sqrt{3}Q)\Bigr)+\frac{1}{2}D_{n}(2Q)\Biggr ]  \nonumber \\
&+&3\Biggl [\sum_{n}\frac{\xi _{n}}{\Bigl[\xi _{n}^{2}+\gamma ^{2}(Q)\Bigr%
]^{2}}\Biggr ]^{2}\Bigl[\sum_{n}\xi _{n}^{-2}\Bigr]^{-1}\Biggr \},
\label{a4}
\end{eqnarray}
\end{widetext}
with
\begin{equation}
D_{n}(Q)=\frac{\xi _{n}^{2}-\gamma ^{2}(Q)}{\xi _{n}^{2}+\gamma ^{2}(Q)}
\label{D}
\end{equation}


\subsection{\label{MFPD.FECDWS.FEUCDWS}Free energy of the unidirectional CDW state}


The CDW order parameter of the unidirectional state (stripe phase)
is~\cite {FPA,KFS,MC}
\begin{equation}
\Delta (\mathop{\bf q}\nolimits)=\frac{(2\pi )^{2}}{L_{x}L_{y}}\Delta (Q)%
\Bigl[\delta (\mathop{\bf q}\nolimits-\mathop{\bf Q}\nolimits)+\delta (%
\mathop{\bf q}\nolimits-\mathop{\bf Q}\nolimits)\Bigr],  \label{DeltaU}
\end{equation}
where the vector $\mathop{\bf Q}\nolimits$ is oriented along the
spontaneously chosen direction, and from Eqs.(\ref{so2}),(\ref{to2}),(\ref
{fo2}) and (\ref{FE4}), the free energy of the unidirectional CDW state
reads
\begin{equation}
{\cal F}^{u}={\cal F}_{0}+4 \frac{L_{x}L_{y}}{2\pi
l_{H}^{2}}T_{0}(Q) \Bigl[b_{2}\Delta ^{2}+b_{4}\Delta ^{4}\Bigr].
\label{FEUni}
\end{equation}
Here the coefficients $b_{2}$ and $b_{4}$ of the Landau expansion are
\begin{equation}
b_{2}=\frac{a_{2}}{3}=\Bigl[1-\frac{T_{0}(Q)}{\pi ^{2}T}\sum_{n}\frac{1}{\xi
_{n}^{2}+\gamma ^{2}(Q)}\Bigr]  \label{b2}
\end{equation}
and
\begin{widetext}
\begin{equation}
b_{4}=\frac{4T_{0}^{3}(Q)}{\pi ^{4}T^{3}}\Biggl \{\sum_{n}\frac{\xi _{n}^{4}%
}{\Bigl[\xi _{n}^{2}+\gamma ^{2}(Q)\Bigr]^{4}}\Biggl [D_{n}(0)+\frac{1}{2}%
D_{n}(2Q)\Biggr ]+2\Biggl [\sum_{n}\frac{\xi _{n}}{\Bigl[\xi _{n}^{2}+\gamma
^{2}(Q)\Bigr]^{2}}\Biggr ]^{2}\Bigl[\sum_{n}\xi _{n}^{-2}\Bigr]^{-1}\Biggr \}%
.  \label{b4}
\end{equation}
\end{widetext}

We note that in the limit $1/\tau \to 0$ evaluation of frequency
sums in expressions for free energy of the triangular and
unidirectional CDW states (\ref{FETri})-(\ref{b4}) leads to the
results obtained in Refs.~\cite{FPA,MC} in the clean case.


\section{\label{MFPD}Mean-field phase diagram}



\subsection{\label{MFPD.IL}Instability line}


The vanishing of the coefficient in front of the quadratic term in the
Landau expansion of the free energy signals about the instability of the
normal state towards the formation of the CDW. This instability corresponds
to the second order phase transition from the homogeneous state to the CDW
state. As usual, the specific parameters of the forming CDW state are
determined by the high order terms in the Landau expansion.

From Eqs. (\ref{a2}) and (\ref{b2}) we obtain the following
equation
\begin{equation}
\frac{T}{T_{0}(Q)}=\frac{1}{\pi ^{2}}\sum_{n}\frac{1}{\xi _{n}^{2}+\gamma
^{2}(Q)}  \label{spin0}
\end{equation}
for the instability line. The solution $T(Q)$ of this equation depends on
the modulus $Q$ of \ the vector that characterizes the CDW state. The
temperature $T_{2}$ of the second order phase transition corresponds to the
maximal value of $T(Q)$:
\begin{equation}
T_{2}=
\mathrel{\mathop{\max }\limits_{Q}}%
T(Q)
\label{T2}
\end{equation}
and the corresponding value $Q_{0}$, $T_{2}=T(Q_{0})$, determines
the period of the CDW state. The Hartree-Fock potential
(\ref{UHF}) has minima at those vectors $Q_{k}$ for which the
form-factor $F_{N}(Q_{k})$ vanishes. In the
clean case this corresponds to $Q_{0}=\min Q_{k}=r_{0}/R_{c}$, where $%
r_{0}\approx 2.4$ is the first zero of the Bessel function of the first kind~
\cite{KFS}. It can be seen from Eq.(\ref{spin0}) that a weak disorder does
not shift the vector $Q_{0}$ (see Appendix). Thus the equation for the
temperature of the second order phase transition into the CDW state reads
\begin{equation}
\frac{T}{T_{0}}=\frac{2}{\pi ^{2}} \Re \psi ^{^{\prime }}\left(
\frac{1}{2}+\frac{1}{4\pi T\tau }+i\,\frac{\mu } {2\pi T}\right)
\label{spin}
\end{equation}
where $\psi ^{^{\prime }}(z)$ is the derivative of digamma
function, $\Re$ the real part, and $T_{0}\equiv T_{0}(Q_{0})$ the
temperature of the transition in the clean case.

Eq.(\ref{spin}) contains the chemical potential $\mu $ that, together with
the temperature $T$ and the broadening of Landau levels $1/2\tau $,
determines the filling factor $\nu _{N}=\nu -2N$ of the partially filled
highest Landau level. However, in order to find this relation one needs to
know the density of states in the system. This question about the density of
states is a very subtle~\cite{KMT} and beyond the scope of the present
paper. For this we use the chemical potential $\mu $ rather than filling
factor $\nu _{N}$.

Eq.(\ref{spin}) can be solved analytically in the two extreme cases: when
temperature $T$ is closed to the temperature $T_{0}$ of instability in the
absence of disorder, and when the temperature $T$ is close to zero.

In the first case, the broadening of the Landau level $1/2\tau $ and the
chemical potential $\mu $ are small compared to the temperature $T_{0}$ of
the instability in the clean case, and, therefore, the leading order
expansion in powers of $1/T_{0}\tau $ and $\mu /T_{0}$ is legitimate. It
appears that the presence of disorder decreases the temperature of
instability linearly:
\begin{equation}
\frac{T}{T_{0}}=1-\frac{7\zeta (3)}{\pi ^{3}T_{0}\tau }-\frac{\mu ^{2}}{%
4T_{0}^{2}}\qquad ,\qquad \frac{1}{2\tau },\mu \ll 2\pi T.  \label{smd}
\end{equation}

In the opposite case $T\rightarrow 0$, one has $1/2\tau ,\mu \gg 2\pi T$,
and Eq.(\ref{spin}) reduces to
\begin{equation}
\frac{\pi }{8T_{0}\tau }=\frac{1}{1+4\mu ^{2}\tau ^{2}}\left[ 1-\Bigl%
(1-12\mu ^{2}\tau ^{2}\Bigr)\frac{\pi ^{4}T^{2}}{48T_{0}^{2}}\right] .
\label{lad}
\end{equation}
We see from Eq.(\ref{lad}) that at zero temperature the second order phase
transition can occur only when the broadening of the Landau level is smaller
than some critical value, $1/2\tau \leq 1/2\tau _{c}=4T_{0}/\pi $.

\begin{figure}
\includegraphics[width=240pt,height=220pt]{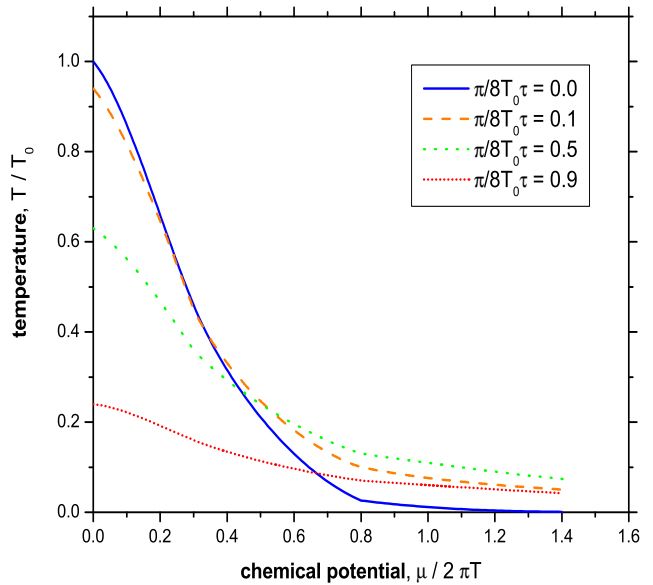}
\caption{ The spinodal lines obtained from Eq.(\ref{spin}) are shown for
different values of dimensionless parameter $\protect\pi /8T_{0}\protect\tau
$.}
\label{Fig.MFPD.IL}
\end{figure}

For other cases Eq.(\ref{spin}) can be solved numerically, and the
corresponding instability (spinodal) line is shown in Fig.~\ref{Fig.MFPD.IL}.
\begin{figure}
\includegraphics[width=240pt,height=220pt]{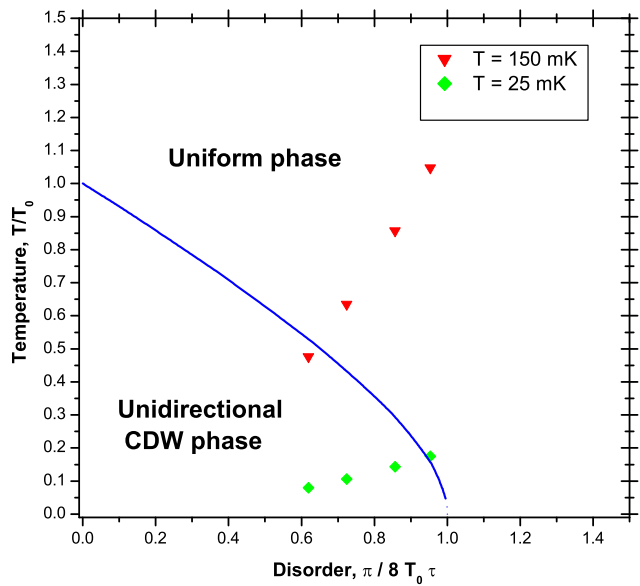}
\caption{ Phase diagram at $\protect\nu _{N}=1/2$. The spinodal
line obtained from Eq.(\ref{tran}) is shown by the solid line. The
triangles and rhombi are the experimental data after Ref.
~\protect\cite{LCEPW}. The Landau level index $N=2,3,4,5$
increases from left to right.} \label{FIG.MFPD.HFLL}
\end{figure}


\subsection{\label{MFPD.HFLL}Half-filled Landau level ($\protect\nu _{N}=1/2$)}


We now consider the case of the half-filled $N$th Landau level ($\nu _{N}=1/2
$), that is related to the recent experiments~\cite{LCEPW}. In this case the
chemical potential is zero, $\mu =0$, provided the density of states is
symmetric around the center of the $N$th Landau level. As follows from Eq.(%
\ref{spin}), the temperature of the second order phase transition for this
case can be found from the equation
\begin{equation}
\frac{T}{T_{0}}=\frac{2}{\pi ^{2}}\zeta \left( 2,\frac{1}{2}+\frac{1}{4\pi
T\tau }\right) ,  \label{tran}
\end{equation}
where $\zeta (2,z)=\sum_{m=0}^{\infty }(m+z)^{-2}$ is the generalized
Riemann zeta function. The analytical solutions of this equation in the
cases of high and low temperature can be obtained from Eqs.(\ref{smd}) and (%
\ref{lad}) by putting $\mu $ to zero. The entire behavior of the spinodal
line, obtained numerically from Eq.(\ref{tran}), is shown in Fig.~\ref
{FIG.MFPD.HFLL}.

We mention that at $\nu _{N}=1/2$ the coefficient $a_{3}$ vanishes
due to the particle-hole symmetry. It means that the transition
from the normal state into the CDW state is of the second order
for both cases of unidirectional and triangular lattice symmetry.
Therefore, to find the structure of the CDW state, one has to take
into account the fourth order terms in the the Landau expansion.
In the vicinity of the spinodal line, it follows from
Eqs.(\ref{a4}) and (\ref{b4}) with $\mu =0$ that
\begin{equation}
a_{4}=\frac{12T_{0}^{3}}{\pi ^{4}T^{3}}\Bigl[-7\zeta (4,u)+12\Phi
_{0}(u)+6\Phi _{2}(u)+8\Phi _{\sqrt{3}}(u)\Bigr]  \label{a412}
\end{equation}
and
\begin{equation}
b_{4}=\frac{2T_{0}^{3}}{\pi ^{4}T^{3}}\Bigl[-3\zeta (4,u)+4\Phi
_{0}(u)+2\Phi _{2}(u)\Bigr],  \label{b412}
\end{equation}
where we introduce the new variable $u=1/2+1/4\pi T\tau $ and the new
function
\begin{eqnarray}
\Phi _{a}\left( \frac{1}{2}+z\right)  &=&\frac{1}{z^{2}{\cal J}_{0}^{2}(ar_{0})}\Biggl [\zeta \left( 2,\frac{1}{2}+z\right)  \\
&-&\frac{1}{z{\cal J}_{0}(ar_{0})}\Im \psi \left(
\frac{1}{2}+z+i\,z{\cal J}_{0}(ar_{0})\right) \Biggr ]  \nonumber
\label{Phi}
\end{eqnarray}
with $\Im$ being the imaginary part. With these expressions we minimized ${\cal F}%
^{t,u}$ with respect to the order parameter $\Delta $ and found that the
unidirectional CDW state has lower free energy.


\begin{figure}
\includegraphics[width=240pt,height=220pt]{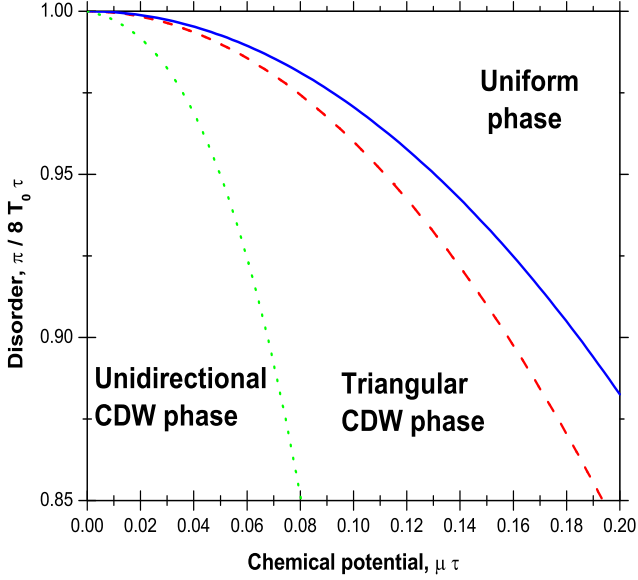}
\caption{Phase diagram at zero temperature near $\protect\nu _{N}=1/2$. The
solid line is obtained from Eq.(\ref{eta1}), the dashes are the spinodal
line and the dots are obtained from Eq.(\ref{eta2}).}
\label{FIG.MFPD.PDZT}
\end{figure}


\subsection{\label{MFPD.PDZT}Phase diagram at zero temperature}


In this section we analyze the zero temperature phase diagram in the case
where the Landau level broadeding is close to its critical value  $1/2\tau
_{c}=4T_{0}/\pi $. Under these conditions, the CDW order parameter $\Delta $
is small, and one can use the Landau expansions (\ref{FETri}) and (\ref
{FEUni}) at zero temperature. The coefficients of these expansions are
\begin{equation}
a_{2}=3\Bigl(1-\frac{8T_{0}\tau }{\pi }H_{1}(\mu \tau )\Bigr
)\,\,,\,\,a_{3}=2\pi \left( \frac{8T_{0}\tau }{\pi }\right)
^{2}H_{2}(\mu \tau ),  \label{a230}
\end{equation}
and
\begin{equation}
a_{4}=3\pi ^{2}\left( \frac{8T_{0}\tau }{\pi }\right) ^{3}H_{3}(\mu \tau
,QR_{c})  \label{a40}
\end{equation}
for the triangular CDW state, and
\begin{equation}
b_{2}=\frac{a_{2}}{3}\,\,,\,\,b_{4}=\frac{\pi ^{2}}{2}\left( \frac{%
8T_{0}\tau }{\pi }\right) ^{3}H_{4}(\mu \tau ,QR_{c})  \label{b0}
\end{equation}
for the unidirectional CDW state. Here we introduce four functions $H_{i}(z)$
as
\begin{equation}
H_{1}(z)=\frac{1}{1+4z^{2}}\,\,,\,\,H_{2}(z)=4zH_{1}(z),  \label{H12}
\end{equation}
\begin{eqnarray}
H_{3}(z,r)
&=&\frac{1}{2}\frac{\displaystyle108z^{2}-5}{(1+4z^{2})^{3}}
+3R_{0}(z,r)+2R_{1}(z,r)  \nonumber  \label{H3} \\
&+&\frac{1}{2}R_{2}(z,r)+2R_{\sqrt{3}}(z,r),
\end{eqnarray}
and
\begin{equation}
H_{4}(z,r)=\frac{\displaystyle28z^{2}-1}{(1+4z^{2})^{3}}%
+2R_{0}(z,r)+R_{2}(z,r),  \label{H4}
\end{equation}
where
\begin{equation}
R_{a}(z,r)=\frac{2H_{1}(z)}{{\cal J}_{0}^{2}(ar)}-\frac{1}{{\cal
J}_{0}^{3}(ar)}\arctan \frac{2{\cal J}_{0}(ar)}{1+4z^{2}-{\cal
J}_{0}^{2}(ar)}. \label{R}
\end{equation}

These expressions result in the following equation on the line of the first
order transition from the uniform to the triangular CDW state
\begin{equation}
\frac{\pi }{8T_{0}\tau }=H_{1}(\mu \tau )+\frac{H_{2}^{2}(\mu \tau )}{%
9H_{3}(\mu \tau ,QR_{c})}.  \label{eta1}
\end{equation}
As before, the maximum of the solutions $1/\tau(Q)$ of Eq.(\ref{eta1}) with
respect to $Q$ should be found. It appears that the maximum is not exactly
at $Q=Q_{0}$ as in the clean case \cite{MC}, but at some shifted value $%
Q_{0}+\delta Q$ with the shift $\delta Q=-0.02(\mu \tau
)^{2}R_{c}^{-1}$ for small $\mu \tau \ll 1$. The existence of the
shift is a feature of the disordered case. Below in the limit $\mu
\tau \ll 1$, we will neglect this shift. In this case
Eq.(\ref{eta1}) can be written as
\begin{equation}
\frac{\pi }{8T_{0}\tau }=1-2.94(\mu \tau )^{2}.  \label{eta11}
\end{equation}

By comparing the free energies of the triangular and the unidirectional CDW
states, we can find the line of the first order transition between them
\begin{equation}
\frac{\pi }{8T_{0}\tau }=H_{1}-\frac{H_{2}^{2}[2H_{3}+H_{4}][3H_{4}+\sqrt{%
H_{4}^{2}+2H_{3}H_{4}}]}{2H_{3}(2H_{3}-3H_{4})^{2}},  \label{eta2}
\end{equation}
For the case $\mu \tau \ll 1$\ Eq.(\ref{eta2}) can be simplified as
\begin{equation}
\frac{\pi }{8T_{0}\tau }=1-18.44(\mu \tau )^{2}.  \label{eta22}
\end{equation}

For other values of $\mu \tau $ Eqs.(\ref{eta1}) and
Eq.(\ref{eta2}) were solved numerically, and the results are shown
in Fig.~\ref{FIG.MFPD.PDZT}. When the parameter $8/\pi T_{0}\tau $
decreases at a fixed value of the chemical potential, the CDW
order parameter grows, and hence, we go beyond the applicability
of to the Landau expansion.

\section{\label{WCC}Weak crystallization corrections}


The CDW order parameter $\Delta(\vvr)$ introduced in
Eq.(\ref{rho}) can be thought of as a saddle-point solution for
the plasmon field that appears in the Hubbard-Stratonovich
transformation of the electron-electron interaction in the action
(\ref{Sinit}). The Landau expansions (\ref{FETri}) and
(\ref{FEUni}) for the free energy of the CDW states were derived
under the assumption that one can neglect the fluctuations of the
CDW order parameter. This is legitimate for $N \gg 1$ and not very
close to the transition (outside the critical region). However,
when one approaches the instability line, the fluctuations of the
CDW order parameter increase. To analyse the effects of the order
parameter fluctuation, we introduce, following the original ideas
of Brazovski~\cite{Braz}, the fluctuations of the CDW order
parameter $\Delta (\mathop{\bf r}\nolimits) \to \Delta
(\mathop{\bf r}\nolimits)+\delta (\mathop{\bf r}\nolimits)$ in the
Landau expansion of the free energy and average over the
fluctuations $\delta (\mathop{\bf r}\nolimits)$. We present below
the results of the corresponding analysis only for the most
interesting case of the half-filled Landau level.

We find that the transition from the uniform to the unidirectional
CDW state becomes of the first order, and takes place at the lower
temperature that can be found from the following equation  (see
Eq.(\ref{tran} for comparison )
\begin{equation}
 \frac{T}{T_{0}} = \frac{2}{\pi^{2}}\zeta\left(2,\frac{1}{2}+ \frac{1}{4\pi T\tau}\right
 )
-g\left (\frac{1}{4\pi T\tau}\right ) N^{-2/3}, \label{TempShift}
\end{equation}
Here function $g(z)$ is defined as
\begin{equation}
g(z)=3\left[ \frac{3\pi r_{0}}{16} \right]^{2/3}\left [ \frac{
\lambda_{0}^{2}(z)}{f(z)}\right ]^{2/3}\left [ \frac{2\lambda
_{0}(z)+\lambda _{2}(z)}{ 4\lambda _{0}(z)-\lambda _{2}(z)}\right]
^{1/3} \label{g}
\end{equation}
where we introduce the following three functions
\begin{equation}
f(z)=\frac{2}{\pi ^{2}}\Biggl [\beta _{1}\zeta \left(
2,\frac{1}{2} +z\right) +\beta _{2}z^{2}\zeta \left(
4,\frac{1}{2}+z\right) \Biggr ], \label{f1}
\end{equation}
\begin{equation}
\lambda_{a}(z)=\frac{2}{\pi ^{4}}\Biggl [-\zeta \left(
4,\frac{1}{2 }+z\right) +2\Phi_{a}\left( 4,\frac{1}{2
}+z\right)\Biggr ], \label{l}
\end{equation}
The constants $\beta_{i}$ are given by
\begin{equation}
\beta _{1}=\frac{T_{0}^{\prime}(Q_{0})}{T_{0}(Q_{0})}\approx
2.58\,\,,\,\,\beta _{2}=\left( {\cal J}_{0}^{\prime
}(Q_{0})\right) ^{2}\approx 0.27,  \label{beta12}
\end{equation}
and function $\Phi_{a}$ is defined by Eq.(\ref{Phi}).

We mention that the function $\pi^{2} g(z)/2 \zeta(2,1/2+z)$
decreases monotonically from the value $0.35$ at $z=0$ to zero at
$z\to \infty$. Therefore, we obtain the following inequality for
the shift $\delta T$ of the mean-field transition temperature $T$
\begin{equation}
\frac{\delta T}{T} \leq 3 \left ( \frac{\pi r_{0}}{16
\sqrt{\beta_{1}}}\right )^{2/3} N^{-2/3}, \qquad \qquad N \gg 1
\label{estimate}
\end{equation}
(the equality corresponds to the clean case.

The appearance of a noninteger powers in Eq.(\ref{TempShift})
results from the fact that the momentum dependence of the
correlation function for the order parameter fluctuations contains
$(Q-Q_{0})^{2}$ rather than $Q^{2}$ (see Ref.~\cite{Braz}).

Eq.(\ref{TempShift}) was derived under the assumption that the
main contribution in the momentum space comes from the region $Q
\approx Q_{0}$. This assumption is justified under the following
condition~\cite{Braz}
\begin{equation}
\frac{g\left (\displaystyle \frac{1}{4\pi T\tau}\right
)}{r_{0}^{2} f\left (\displaystyle \frac{1}{4\pi T\tau}\right )}
\ll N^{2/3} \label{validity}
\end{equation}
The combination of functions in the left hand side of inequality
(\ref{validity}) decreases monotonically from $0.023$ to $0$ while
$z$ increases from zero to infinity and, hence, the condition
(\ref{validity}) is hold.

According to Eq.(\ref{estimate}), the fluctuations reduce the
transition temperature by the amount of the order of $N^{-2/3}\ll
1$ and, therefore, in the considered case of the weak magnetic
field ($N \gg 1)$ their effects can be neglected. These results
indicate that the critical region for the considered transition is
indeed small, and the mean-field approach gives a good
approximation for $N \gg 1$.

\section{\label{Disc}Discussions}


\subsection{\label{Disc.E}Comparison with experimental results}


Now we discuss the possible applications of our theory to the
recent experiments. Although our mean-field theory was derived for
the case of a large number of the occupied Landau level $N\gg 1$,
and neglects corrections of the order of $1/N$, while
experimentally one has $N=2,3,4$,  we however expect that
Eq.(\ref{spin}) gives a good estimation for the temperature of the
transition from the uniform to the CDW state, even for $N=2,3,4$.
We have complementary assurance that it can really be the case
because Eq.(\ref{spin}) can be obtained without introducing the
CDW order parameter and considering the mean-field theory but as
the equation that determines the temperature $T(Q)$ at which the
two-particle vertex function at wave vector $Q$
diverges~\cite{Vertex}.

We restrict ourselves by discussion of the experiments without an
in-plane magnetic field~\cite{LCEPW,DTSPW}. The theory for the
half-filled highest Landau level contains two physical parameters:
the temperature $T_{0}$ and the broadening of the Landau level
$1/2\tau $. The estimate of the temperature of instability $T_{0}$
in the absence of disorder is more subtle. The theory can provide
an estimate for $U(Q_{0})$ and, correspondingly, for the value
of $T_{0}$ only for a weak magnetic field. In this case the value of $%
U(Q_{0})$ can be found analytically~\cite{KFS} with the following
result for $T_{0}$
\begin{equation}
T_{0}=\frac{\alpha }{4}\omega _{H}\qquad ,\qquad Nr_{s}\gg 1\,,\,r_{s}\leq 1,
\label{alpha}
\end{equation}
where the numerical parameter $\alpha $ only slightly depends on $r_{s}$ in
the range $0.1<r_{s}<1$, and approximately $\alpha \approx 0.03$.
The broadening of the Landau level $1/2 \tau$ can be estimated from the
mobility at zero magnetic field. However, the results obtained in
Sec.~\ref{MFPD.HFLL} impose the restriction on the value of the sample
mobility at zero magnetic field. In order to observe the CDW states at the
partially filled Landau level with index $N$ the mobility at zero magnetic
field should satisfy the condition $\mu > e N / 2 \alpha^{2} n_{e}$
where $n_{e}$ denotes the electron density of 2DEG. For typical
values of the electron
density of 2DEG we can obtain the following estimate $\mu > N \cdot 10^{6}
cm^{2} /V s$.

In the experiments of Lilly et al.~\cite{LCEPW}, the samples were
relatively clean (mobilities exceed $9\cdot 10^{6}\,cm^{2}/Vs$),
and the anisotropy was observed for $N=2$ below $T=150\,mK$. For
the temperature $T=25\,mK$, the anisotropy in the resistance was
found to disappear with the decreasing the magnetic field for
$N=4$. We plot experimental points in Fig.~\ref {FIG.MFPD.HFLL}
under the assumption that Eq.(\ref{alpha}) remains valid even for
$N=2,3,4,$ and $5$. As it can be seen from this Figure, for a
constant magnetic field the highest temperature at which
anisotropy appears, corresponds to $N=2$. As the magnetic field
decreases at some constant temperature, the anisotropy disappears
at some value of $N$ due to the disorder induced transition from
the CDW state to the uniform. It should be mentioned that without
the effects of disorder the anisotropy in the resistance should
remains up to $N=12$ at $T=25mK$. Therefore, the disorder plays an
important role even in the high mobility samples.


\subsection{\label{Disc.N}Comparison with numerical results}


The problem of the formation of the CDW state on the second Landau level
with $\nu _{N}=1/2$ at zero temperature in the presence of a quenched
disorder was studied numerically in Ref.~\cite{SWF}. The system of 12 electrons
interacting via the Coulomb interaction $U(q) = 2 \pi e^{2} /q$
in the presence of the
quenched disorder was projected on the second Landau level ($N=2$). The effects
of interactions with electrons on the other Landau level was not taken into
account. The system was diagonalized numerically.
It was found that the CDW state transforms into a
uniform liquid state as the dimensionless disorder strength
$\omega_{H} \sqrt{n_{L}/2\pi \rho \tau_{0}}$ exceeds $0.12$.

In order to be able to compare the results of the presented above mean-field
theory with the numerical results, we perform the evaluation of
the temperature $T_{0}$ in the case for which the numerical results were
obtained (instead screened interaction (\ref{U0}) we use
$U(q) = 2 \pi e^{2} /q$). Under this curcumstances our theory gives
the value $0.14$.

The small discrepancy may be attributed to two factors: on the one
hand, the finite number of electrons in numerical calculations
and, on the one hand, unsufficience of the Hartree-Fock
approximation for the problem with the Coulomb interaction $U(q) =
2 \pi e^{2} /q$. In the later case one should take into account
the diagrams beyond the Hartree-Fock theory. Nevertheless, the
comparison demonstrates that such corrections are small.

We emphasize that our theory which takes into account the screening of
electron-electron interaction by electrons on the other Landau levels
gives much smaller value 0.038 of the dimensionless disorder strength
for the transition from the uniform liquid state to the CDW state.


\section{\label{Conc}Conclusions}


For the system of a two-dimensional interacting electrons in the
presence of a weak disorder and a weak magnetic field, we
investigated the effect of disorder on the existence of the
mean-field CDW states in the framework of the Hartree-Fock
approximation. In the considered case of large filling factors
$\nu \gg 1$, we obtained that the mean-field CDW instability
exists if the disorder is rather weak, $1/\tau \leq 8T_{0}/\pi $.
We found that at half-filling the unidirectional CDW state
appears, and the presence of disorder does not change the vector
of the CDW. Near half-filling, the unidirectional CDW state is
energetically more favorable than the triangular one. We obtained
that the weak crystallization corrections to the mean-field result
are of the order of $(1/\nu)^{2/3} \ll 1$ and thus can be
neglected. We discussed the applications of our theory to the
recent experimental and numerical results.

\acknowledgements
 I.B. is grateful to M.V. Feigel'man and M.A.
Skvortsov for stimulating discussions. This research was supported
by Forschungszentrum J\"ulich (Landau Scholarship), Russian
Foundation for Basic Research (RFBR), and Deutsche
Forschungsgemeinschaft (DFG).


\appendix

\section{\label{App}Instability vector}


In this appendix we prove that the weak disorder does not change
the vector at which the instability towards the formation of the
CDW state grows. Let us consider the solution $T+\delta T$ of
Eq.(\ref{spin0}) for the vector  $Q=Q_{0}+\delta Q$, where $\delta
Q\ll Q_{0}$. We will now show that the shift $\delta T$ is always
negative, and hence, the maximal instability temperature
corresponds to the vector $Q=Q_{0}$, as it is in the clean case.

For a small deviation $\delta Q$ we can write
\begin{equation}
T_{0}(Q)=T_{0}(1-\beta _{1}(\delta QR_{c})^{2})\,\,,\,\,{\cal J}%
_{0}^{2}(QR_{c})=\beta _{2}(\delta QR_{c})^{2}.  \label{T,J}
\end{equation}
The shift $\delta T$ results in the substitution
\begin{equation}
\xi _{n}\rightarrow \xi _{n}-\frac{\delta T}{4\pi T^{2}\tau }-i\frac{\delta
T\mu }{2\pi T^{2}},  \label{ksi}
\end{equation}
in Eq.(\ref{spin0}), and we obtain
\begin{equation}
\frac{\delta T}{T}=-(\delta QR_{c})^{2})\frac{\beta _{1}g_{2}+\beta
_{2}z^{2}g_{4}}{g_{2}-2(zg_{3}+y\overline{g_{3}})},  \label{dt}
\end{equation}
where $z=1/4\pi T\tau $ and $y=\mu /2\pi T$. Here we introduce the four
functions $g_{a}(z,y)$ and $\overline{g_{a}(z,y)}$
\begin{equation}
g_{a}(z,y)= \Re \sum_{n=0}^{\infty }\xi
_{n}^{-a}\,\,,\,\,\overline{g_{a}(z,y)}=\Im \sum_{n=0}^{\infty
}\xi _{n}^{-a}.  \label{gi}
\end{equation}
It can be easily seen that the rhs of Eq.(\ref{dt}) is negative for all
possible values of $z$ and $y$.


\end{document}